\documentclass[conference]{IEEEtran}
\IEEEoverridecommandlockouts
\usepackage{cite}
\usepackage{amsmath,amssymb,amsfonts}
\usepackage{algorithmic}
\usepackage{graphicx}
\usepackage{textcomp}
\usepackage{xcolor}

\usepackage[table]{xcolor}
\usepackage{array}
\usepackage{booktabs}

\definecolor{headercolor}{RGB}{255,255,255} % Orange header
\definecolor{lightorange}{RGB}{255,255,255}
\definecolor{darkred}{RGB}{153,0,0}
\definecolor{bluecell}{RGB}{0,153,204}
\definecolor{greencell}{RGB}{0,153,0}

\def\BibTeX{{\rm B\kern-.05em{\sc i\kern-.025em b}\kern-.08em
    T\kern-.1667em\lower.7ex\hbox{E}\kern-.125emX}}
\begin{document}

\title{A framework to evaluate the performance of  Variational Quantum Algorithms}

\makeatletter
\newcommand{\linebreakand}{%
  \end{@IEEEauthorhalign}
  \hfill\mbox{}\par
  \mbox{}\hfill\begin{@IEEEauthorhalign}
}
\makeatother
 
\author{\IEEEauthorblockN{Ernesto Mamedaliev}
\IEEEauthorblockA{\textit{Innovation Hub} \\
\textit{PricewaterhouseCoopers}\\
San Francisco, USA \\
ernesto.mamedaliev.guesinova@pwc.com}
\and
\IEEEauthorblockN{Vladyslav Libov}
\IEEEauthorblockA{\textit{Innovation Hub} \\
\textit{PricewaterhouseCoopers}\\
San Francisco, USA \\
vladyslav.libov@pwc.com}
\and
\IEEEauthorblockN{Albert Nieto-Morales}
\IEEEauthorblockA{\textit{Innovation Hub} \\
\textit{PricewaterhouseCoopers}\\
San Francisco, USA \\
albert.morales@pwc.com}
\linebreakand 
\IEEEauthorblockN{Oskar S\l{}owik}
\IEEEauthorblockA{\textit{Innovation Hub} \\
\textit{PricewaterhouseCoopers}\\
San Francisco, USA \\
oskar.slowik@pwc.com}
\and
\IEEEauthorblockN{Arit Kumar Bishwas}
\IEEEauthorblockA{\textit{Innovation Hub} \\
\textit{PricewaterhouseCoopers}\\
San Francisco, USA \\
arit.kumar.bishwas@pwc.com}
}

\maketitle

\begin{abstract}
Variational Quantum Algorithms (VQAs) are among the most promising approaches for solving combinatorial optimization problems on noisy intermediate-scale quantum (NISQ) devices. However, benchmarking VQAs remains challenging due to their stochastic nature and the absence of standardized performance criteria. In this work, we introduce a general framework for evaluating VQAs applied to Quadratic Unconstrained Binary Optimization (QUBO) problems. The framework combines three complementary metrics—\textit{feasibility}, \textit{quality}, and \textit{reproducibility}—and visualizes performance through a \textit{quality diagram} that captures trade-offs between success probability and resource consumption. We formalize reproducibility using Shannon entropy and define a decision criterion for algorithm selection under resource constraints. As a proof of concept, we apply the framework to multiple VQAs using Conditional Value at Risk (CVaR) based cost functions and varying shot counts on a 16-qubit QUBO instance. Our results illustrate how the proposed methodology enables systematic benchmarking. This work contributes a principled approach for comparing VQAs in the NISQ era and provides a foundation for adaptive algorithm selection in hybrid quantum-classical workflows.
\end{abstract}


\section{Introduction}
Combinatorial optimization problems~\cite{PapadimitriouSteiglitz1998,WolseyNemhauser1999} arise in diverse domains such as finance~\cite{CornuejolsTutuncu2006,MansiniOgryczakSperanza2015}, logistics~\cite{RenLuGuoXieZhangZhang2023,LouatiLahyaniAldaejMellouliNusir2021}, and machine learning~\cite{blum1997selection,AloiseDeshpandeHansenPopat2009,HyafilRivest1976}. Many of these problems can be formulated as \textit{quadratic unconstrained binary optimization (QUBO) problems}~\cite{kochenberger2014unconstrained,glover2018tutorial}, which are generally NP-hard~\cite{garey1979computers}, making them increasingly challenging for classical algorithms as problem size grows. Classical heuristics, such as simulated annealing~\cite{kirkpatrick1983optimization} and genetic algorithms~\cite{holland1975adaptation}, provide approximate solutions, but their performance deteriorates with increasing problem size, motivating the exploration of alternative approaches. Recent works illustrate this trend across quantum annealing and variational methods~\cite{Bishwas2024PortfolioQA,Bishwas2024ParallelQA,Bishwas2024MolecularUnfolding}.

Quantum information processing~\cite{nielsen2002quantum} has emerged as a promising paradigm for tackling such problems. In the current noisy intermediate-scale quantum (NISQ) era~\cite{preskill2018quantum,bharti2022noisy}, hardware limitations—such as short coherence times and imperfect gate fidelities—restrict achievable circuit depth. Variational Quantum Algorithms (VQAs)~\cite{cerezo2021variational,mcclean2016theory} were introduced to exploit these NISQ devices by combining shallow parameterized quantum circuits (ansätze) with classical optimization routines. VQAs suppress noise effects while partially exploring the quantum state space, making them suitable for near-term applications. Crucial hyperparameters influencing solution quality include ansatz design~\cite{du2022quantum}, measurement strategy (number of shots)~\cite{Zhu2024}, and the choice of classical optimizer~\cite{Jones2024}.

Several studies have explored solving real-world problems through QUBO formulations using VQAs~\cite{perez2024variational,schwagerl2024benchmarking}, and further work has examined the impact of ansatz choice (e.g., QAOA~\cite{farhi2014quantum}) and cost function design (e.g., CVaR~\cite{barkoutsos2020improving}). However, clarity on VQA performance for such applications remains limited because the stochastic nature of these variational algorithms is not always treated explicitly, and no universally accepted criterion exists to classify VQAs according to their suitability for solving a given QUBO~\cite{bonet2021performance,vinci2018optimally,wu2024variational}.

In this work, we address these gaps by introducing a general framework for evaluating VQAs applied to QUBO problems. The framework combines three complementary metrics—\textit{feasibility}, \textit{quality}, and \textit{reproducibility}—and visualizes performance through a \textit{quality diagram} that captures trade-offs between success probability and resource consumption. We formalize reproducibility using Shannon entropy and define a decision criterion for algorithm selection under resource constraints. As a proof of concept, we apply the framework to multiple VQAs using CVaR-based cost functions and varying shot counts on a 16-qubit QUBO instance.

This study is organized as follows. Section~II presents the theoretical background and the proposed framework. Section~III applies the framework to a set of proof-of-concept experiments. Section~IV provides a comprehensive discussion of the results, theoretical implications, and limitations, and concludes with directions for future research.

\section{Theoretical background}

In this section, we introduce our theoretical framework for evaluating VQAs applied to QUBO problems. First, we present the formal definition of a QUBO problem and interpret quantum circuits as probability distributions over the binary space. This allows us to characterize quantum circuits within the context of specific QUBO instances. We then parametrize VQAs and their performance, accounting for their stochastic nature, and define \textit{VQA distributions}. Based on these distributions, we introduce measures of feasibility and reproducibility for variational quantum algorithms. Finally, we propose a function to quantify the quality of a VQA’s performance in solving a QUBO problem.

\subsection{Cost distributions}

A QUBO problem can be formally defined as follows:

\vspace{2mm}

\textit{Given a symmetric matrix $Q\in \mathbb{R}^{N\times N}$ and a function $f:\{0,1\}^N \longrightarrow \mathbb{R}$ defined by}
\begin{equation}
    f(x) = x^T Q x, \quad x \in \{0,1\}^N,
\end{equation}
\textit{find an element $x^* \in \{0,1\}^N$ that minimizes $f$, i.e.,}
\begin{equation}
    f(x^*) \le f(x), \quad \forall x \in \{0,1\}^N.
\end{equation}

\noindent The integer $N$ is referred to as the \textit{dimension} of the problem, and it will be equal to the number of qubits we use to design the quantum circuits. 

In gate-based quantum computation, quantum circuits can be interpreted as probability distributions over the binary space $\{0,1\}^N$ via Born’s rule~\cite{nielsen2002quantum}. The QUBO objective function $f(x)$ maps each bitstring to a real number, representing its \textit{cost}. Consequently, each quantum circuit induces a distribution over costs, which we refer to as the \textit{cost distribution}. Statistical quantities derived from these cost distributions are typically used as objective functions in the classical optimization step of VQAs~\cite{cerezo2021variational,barkoutsos2020improving}, making VQA optimization intrinsically stochastic~\cite{kushner2003stochastic,spall2003introduction}. This inherent stochasticity motivates analyzing \textit{VQA performance} through statistical measures.

\subsection{VQA structure}
To analyze the performance of a VQA, we must first define it in terms of its hyperparameters to avoid ambiguity. Following the literature~\cite{cerezo2021variational,mcclean2016theory,perez2024variational}, we formally specify a VQA as:
\begin{equation}
    VQA := (\mathcal{U}, [C, s, \mathcal{A}]),
\end{equation}
where:
\begin{itemize}
    \item $\mathcal{U} \subseteq \{ U(\vec{\theta}) \in SU(2^N) : \vec{\theta} \in \mathbb{R}^k \}$ denotes the \textit{ansatz}, with $U(\vec{\theta})$ representing a quantum circuit and $k$ the number of parameters required to specify each element of the ansatz.
    \item $C : \mathbb{R}^k \longrightarrow \mathbb{R}$ is the \textit{cost function}, which takes as input the parameters defining an element of the ansatz and outputs a statistical quantity derived from the cost distribution of the corresponding quantum circuit.
    \item $s$ is the \textit{number of shots}, determining the sample size of the cost distribution.
    \item $\mathcal{A} : \mathbb{R}^k \longrightarrow \mathbb{R}^k$ denotes the \textit{classical optimization algorithm}.
\end{itemize}

Thus, a VQA can be viewed as a map:
\begin{equation}
    VQA : SU(2^N) \longrightarrow SU(2^N), \quad U(\vec{\theta}_{\text{in}}) \longmapsto U(\vec{\theta}_{\text{out}}),
\end{equation}
which transforms an input quantum circuit into an output circuit according to the chosen cost function, number of shots, and optimization strategy.

According to~\cite{barkoutsos2020improving,ascendingcvar}, two relevant metrics can be used to quantify the performance of a VQA optimization. The first one is the probability of measuring the global minimum, denoted by \(p_{\mathrm{min}}\), which represents the likelihood that the quantum circuit produces an optimal bitstring upon measurement. A high value of \(p_{\mathrm{min}}\) indicates that the circuit effectively concentrates probability mass around optimal solutions. The second metric is the number of quantum circuit evaluated during optimization, \(n_{\mathrm{calls}}\), which serves as a proxy for the computational resources consumed.

If we cap the number of quantum circuits we are evaluating at a maximum value \(n_{\max}\)—representing the upper bound on resources we are willing to allocate—the performance of a VQA can be represented in a normalized two-dimensional diagram:
\[
    \mathcal{D} = [0,1]^2 \subset \mathbb{R}^2,
\]
defined through the following change of variables:
\begin{equation}
    \begin{cases}
        n_{\mathrm{calls}} \longmapsto \dfrac{n_{\mathrm{calls}} - 1}{n_{\max} - 1},\\[6pt]
        p_{\mathrm{min}} \longmapsto 1 - p_{\mathrm{min}}.
    \end{cases}
\end{equation}

The ideal search process corresponds to the point \((0,0)\) in this diagram, representing a situation where the global minimum is reached with probability \(1\) after a single function call. We refer to this representation as the \textit{quality diagram}.

In the quality diagram, each VQA execution is represented as a point defined by \((n_{\mathrm{calls}}, p_{\mathrm{min}})\). Due to the stochastic nature of the algorithm, repeated runs with identical settings yield not a single point but a distribution of points over \(\mathcal{D}\), referred to as the \textit{VQA distribution} (see Figure~1), that we will treat as a probability distribution. Analyzing this distribution enables us to characterize the quality and reproducibility of the VQA optimization process.

\begin{figure}[!t]
    \centerline{\includegraphics[width=0.5\textwidth]{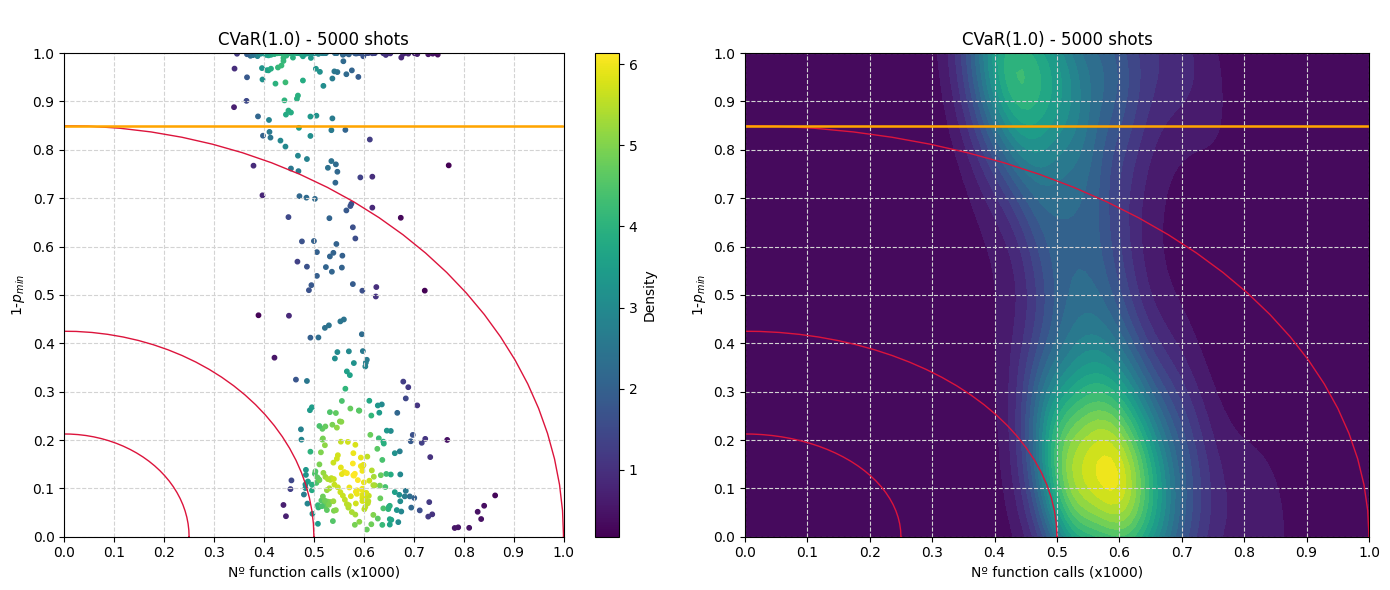}}
    \caption{On the left, the distribution of results obtained by running 400 VQAs to solve a QUBO problem as specified in Section~3. The orange line represents the probability threshold, while the red lines correspond to the level curves of the quality function for \(q = 1\), \(q = 2\), and \(q = 4\) (see Subsection~2.3). On the right, a contour plot is shown for illustrative purposes, representing the underlying distribution of VQA outcomes.}
    \label{fig:distr}
\end{figure}

\subsection{Performance evaluation}

A VQA can be understood as a map that takes a quantum circuit as input and returns a quantum circuit according to a chosen ansatz, cost function, number of measurements, and optimization algorithm. As discussed, a given VQA defines a distribution on the quality diagram \(\mathcal{D}\), so the performance of the variational algorithm should be evaluated based on the characteristics of this distribution. Here, we consider three performance metrics:
\begin{itemize}
    \item \textit{Feasibility}, which measures the a priori probability of succeeding in the task.
    \item \textit{Quality}, which evaluates the efficiency of the circuit search process.
    \item \textit{Reproducibility}, which quantifies the certainty of the results.
\end{itemize}
These three quantities form the basis for a decision criterion to classify the suitability of a VQA for solving a given QUBO problem.

\subsubsection{Feasibility and reproducibility}
Considering a probability threshold to measure a global minimum \(p_{\text{threshold}}\), the feasibility of a VQA is defined as the probability that the output circuit achieves a probability of measuring the global minimum above this threshold:
\begin{equation}
    \mathcal{F}[\rho] = \mathcal{P}(p_{\text{min}} \geq p_{\text{threshold}}),
\end{equation}
where \(\rho\) is the VQA distribution on \(\mathcal{D}\) and \(p_{\text{min}}\) denotes the probability the output circuit has to measure a global minimum. Formally:
\begin{equation}
    \mathcal{F}[\rho] = \int_0^{n_{\text{max}}} \int_{p_{\text{threshold}}}^1 \rho(n_{\text{calls}}, p_{\text{min}}) \, dn_{\text{calls}} \, dp_{\text{min}}.
\end{equation}

Reproducibility can be understood as a measure of the concentration of \(\rho\). In other words, if \(\rho\) is narrow, running the variational algorithm twice is more likely to yield two quantum circuits with the same \(p_{\text{min}}\) using the same amount of resources \(n_{\text{calls}}\). Conversely, a broader VQA distribution implies that the outcomes are more spread across \(\mathcal{D}\). To quantify this spread, we use the Shannon entropy~\cite{shannon1948mathematical} (in its discrete form), since it is maximized for a uniform distribution (maximal spread) and equals zero for a fully concentrated distribution. Therefore, we define reproducibility as:
\begin{equation}
    \mathcal{R}[\rho] = 1 - \frac{1}{S_{\text{max}}} S[\rho],
\end{equation}
where \(S[\rho]\) denotes the Shannon entropy of the VQA distribution and \(S_{\text{max}}\) is its maximum possible value.

\subsubsection{Quality function}

The output of a VQA run is represented as a point in the quality diagram. Given two runs, we aim to quantify which one performs better in terms of the trade-off between achieving a high probability of measuring the global minimum and minimizing the number of quantum circuits evaluated during optimization. To this end, we define a \textit{quality function} that assigns a scalar value to each point in \(\mathcal{D}\). This function must satisfy the following conditions:
\begin{itemize}
    \item If the probability of measuring the global minimum, \(p_{\text{min}}\), is below the probability threshold, \(p_{\text{threshold}}\), the quality of the optimization is zero.
    \item For two optimizations requiring the same number of quantum circuit evaluations, the one with a higher probability of measuring the global minimum has higher quality.
    \item For two optimizations with the same probability of measuring the global minimum, the one requiring fewer quantum circuit evaluations has higher quality.
\end{itemize}

Based on these criteria, we propose the following form for the quality function:
\begin{equation}
    q(n_{\mathrm{calls}}, p_{\mathrm{min}}) = 
    \frac{\sigma(p_{\mathrm{min}} - p_{\mathrm{threshold}})}
    {r(n_{\mathrm{calls}}, p_{\mathrm{min}})},
\end{equation}
where \(\sigma(p_{\mathrm{min}} - p_{\mathrm{threshold}})\) denotes the Heaviside step function (equals to \(1\) if \(p_{\mathrm{min}} > p_{\mathrm{threshold}}\) and \(0\) otherwise), and \(r(n_{\mathrm{calls}}, p_{\mathrm{min}})\) represents a weighted Euclidean distance from the ideal point \((0,0)\) in the quality diagram:
\begin{equation}
    r(n_{\mathrm{calls}}, p_{\mathrm{min}}) =
    \sqrt{
        \left(\frac{n_{\mathrm{calls}} - 1}{n_{\max} - 1}\right)^2 +
        \left(\frac{1 - p_{\mathrm{min}}}{1 - p_{\mathrm{threshold}}}\right)^2
    }.
\end{equation}

Considering this function, we define the quality of a VQA in terms of its distribution as follows:
\begin{equation}
    \mathcal{Q}[\rho] = \iint_{\mathcal{D}} \rho(n_{\text{calls}}, p_{\text{min}}) \, q(n_{\text{calls}}, p_{\text{min}}) \, dn_{\text{calls}} \, dp_{\text{min}},
\end{equation}
where \(\rho\) is the VQA distribution on \(\mathcal{D}\) and \(q(n_{\text{calls}}, p_{\text{min}})\) is the quality function. This quantity captures the stochastic nature of the VQA and evaluates its efficiency in solving a given QUBO problem.

\subsection{Selection criterion}
Given a VQA designed to solve a QUBO problem, we can assign performance metrics that capture the stochastic nature of the algorithm:
\begin{equation}
    VQA := (\mathcal{U}, [\mathcal{C}, s, \mathcal{A}]) \longmapsto (\mathcal{F}[\rho], \mathcal{Q}[\rho], \mathcal{R}[\rho]),
\end{equation}
where \(\mathcal{F}[\rho]\), \(\mathcal{Q}[\rho]\), and \(\mathcal{R}[\rho]\) denote feasibility, quality, and reproducibility, respectively.

To define a selection criterion for considering a VQA as a \textit{good solver}, we address three key questions:
\begin{itemize}
    \item \textit{(i) What is the probability of succeeding in the task?} Feasibility provides this a priori probability.
    \item \textit{(ii) How resource-efficient is the algorithm?} Quality answers this by capturing the trade-off between the probability of sampling the global minimum and the number of circuits evaluated during optimization.
    \item \textit{(iii) How confident are we in the feasibility and quality metrics obtained?} Reproducibility quantifies how reliable the results are by considering the spread of the VQA distribution.
\end{itemize} We propose defining performance metric thresholds \((\mathcal{F}_0, \mathcal{Q}_0, \mathcal{R}_0)\) such that the selection process works as follows. Given a VQA and its VQA distribution \(\rho\), we can compute its performance metrics \((\mathcal{F}[\rho], \mathcal{Q}[\rho], \mathcal{R}[\rho])\), so:
\begin{enumerate}
    \item If \(\mathcal{F}[\rho] < \mathcal{F}_0\), discard the VQA as a solver; otherwise, proceed to the next step.
    \item If \(\mathcal{Q}[\rho] < \mathcal{Q}_0\), discard the VQA as a solver; otherwise, proceed to the next step.
    \item If \(\mathcal{R}[\rho] < \mathcal{R}_0\), discard the VQA as a solver; otherwise, consider it a well-suited solver.
\end{enumerate}

This approach enables us to classify VQAs according to their performance in solving QUBO problems. In the next section, we present a specific QUBO instance and evaluate multiple VQAs to solve it. This selection criterion allows us to identify the best-performing algorithms.

\section{Experimental procedure}
In this section, we present a series of numerical experiments, conducted through simulation, as a proof of concept for the theoretical framework introduced earlier. The goal is to evaluate several VQAs in solving a fixed random QUBO problem of dimension \(N = 16\) using the previously defined selection criterion. The QUBO instance remains the same across all experiments (available upon reasonable request).

Additionally, we aim to analyze the effect of stochasticity on the results and observe the influence of varying the number of shots and the CVaR parameter \(\alpha\). All optimizations start from the same randomly chosen initial quantum circuit (available upon request).

\subsection{VQA instances}
Given the VQA structure \( VQA := (\mathcal{U}, [\mathcal{C}, s, \mathcal{A}]) \), we specify its parameters as follows:
\begin{itemize}
    \item \textbf{Ansatz:} \(\mathcal{U} \subseteq SU(2^{16})\) is given by the \textit{RealAmplitudes} ansatz~\cite{he2022,abbas2021,arthur2022}, which is the default ansatz in Qiskit’s variational quantum circuit implementation. We use the reverse-linear entanglement scheme with one repetition.
    \item \textbf{Cost functions:} We consider several cost functions \( C_\alpha : \mathbb{R}^{32} \longrightarrow \mathbb{R} \) based on CVaR~\cite{barkoutsos2020improving,ascendingcvar}. The values of \(\alpha\) considered are \(\{0.15, 0.25, 0.50, 0.75, 1.00\}\).
    \item \textbf{Number of shots:} \( s \in \mathbb{N} \), defining the sample size of the cost distribution, is studied in the range \(\{1\,000, 5\,000, 10\,000, 20\,000, 30\,000, 40\,000, 60\,000,\\
    80\,000, 100\,000\}\).
    \item \textbf{Optimizer:} The parameter optimization algorithm \(\mathcal{A} : \mathbb{R}^{32} \longrightarrow \mathbb{R}^{32}\) is given by the COBYLA optimizer~\cite{powell1994cobyla}.
\end{itemize}

To ensure statistical consistency in the results, we run \(n = 400\) optimizations for each VQA instance (see Appendix~\ref{app:qubo-matrix} for further details).

\subsection{VQA selection}
To identify which VQAs perform better on the given QUBO instance, we conduct three tests following the steps outlined in Subsection~II.D.
\subsubsection{Step 1 (Feasibility test)}
For feasibility, we consider \(\mathcal{F}_0 = 0.70\), meaning that we require an a priori probability of at least \(0.70\) for success. Table~1 presents the feasibility results obtained for each VQA with a 95\% confidence level. VQAs that pass the feasibility test are highlighted in red. Additionally, these results are plotted in Figure~2 to illustrate how feasibility scales with the number of shots for each cost function considered. The results show that feasibility increases with both the value of \(\alpha\) and the number of shots \(s\), until reaching saturation—indicating that further increasing the number of shots has negligible impact on feasibility.
\begin{table}[!t]
\caption{Feasibilities of VQAs, for different $s$ and $\alpha$ values, with a confidence level of $95\%$.}
\centering
\renewcommand{\arraystretch}{1.3}
\setlength{\tabcolsep}{2pt}
\begin{tabular}{|>{\bfseries}c| *{5}{c|}}
\hline
\rowcolor{headercolor!50}
\boldmath $s\backslash\alpha$ & \bfseries 0.15 & \bfseries 0.25 &\bfseries 0.50 &\bfseries 0.75 &\bfseries 1.00 \\
\hline
\cellcolor{headercolor!50}\bfseries 1\,000 & 0.16 $\pm$ 0.04 & 0.23 $\pm$ 0.05 & 0.37 $\pm$ 0.05 & 0.34 $\pm$ 0.05 & 0.15 $\pm$ 0.04 \\
\cellcolor{headercolor!50}\bfseries 5\,000 & 0.35 $\pm$ 0.05 & 0.31 $\pm$ 0.05 & 0.56 $\pm$ 0.05 & \cellcolor{darkred!20}{0.70 $\pm$ 0.05} & \cellcolor{darkred!20}{0.69 $\pm$ 0.05} \\
\cellcolor{headercolor!50}\bfseries 10\,000 & 0.46 $\pm$ 0.05 & 0.24 $\pm$ 0.05 & 0.60 $\pm$ 0.05 & \cellcolor{darkred!20}{0.75 $\pm$ 0.05} & \cellcolor{darkred!20}{0.78 $\pm$ 0.04} \\
\cellcolor{headercolor!50}\bfseries 20\,000 & 0.40 $\pm$ 0.05 & 0.25 $\pm$ 0.05 & \cellcolor{darkred!20}{0.70 $\pm$ 0.05} & \cellcolor{darkred!20}{0.69 $\pm$ 0.05} & \cellcolor{darkred!20}{0.87 $\pm$ 0.04} \\
\cellcolor{headercolor!50}\bfseries 30\,000 & 0.40 $\pm$ 0.05 & 0.30 $\pm$ 0.05 & \cellcolor{darkred!20}{0.71 $\pm$ 0.05} & \cellcolor{darkred!20}{0.70 $\pm$ 0.05} & \cellcolor{darkred!20}{0.92 $\pm$ 0.03} \\
\cellcolor{headercolor!50}\bfseries 40\,000 & 0.41 $\pm$ 0.05 & 0.40 $\pm$ 0.05 & \cellcolor{darkred!20}{0.69 $\pm$ 0.05} & {0.65 $\pm$ 0.05} & \cellcolor{darkred!20}{0.90 $\pm$ 0.03} \\
\cellcolor{headercolor!50}\bfseries 60\,000 & 0.41 $\pm$ 0.05 & 0.50 $\pm$ 0.05 & \cellcolor{darkred!20}{0.71 $\pm$ 0.05} & \cellcolor{darkred!20}{0.70 $\pm$ 0.05} & \cellcolor{darkred!20}{0.95 $\pm$ 0.03} \\
\cellcolor{headercolor!50}\bfseries 80\,000 & 0.38 $\pm$ 0.05 & 0.55 $\pm$ 0.05 & \cellcolor{darkred!20}{0.74 $\pm$ 0.05} & \cellcolor{darkred!20}{0.72 $\pm$ 0.05} & \cellcolor{darkred!20}{0.93 $\pm$ 0.03} \\
\cellcolor{headercolor!50}\bfseries 100\,000 & 0.40 $\pm$ 0.05 & 0.55 $\pm$ 0.05 & \cellcolor{darkred!20}{0.75 $\pm$ 0.05} & \cellcolor{darkred!20}{0.76 $\pm$ 0.05} & \cellcolor{darkred!20}{0.95 $\pm$ 0.03} \\
\hline
\end{tabular}

\end{table}

\begin{figure}[!t]
    \centerline{\includegraphics[width=0.5\textwidth]{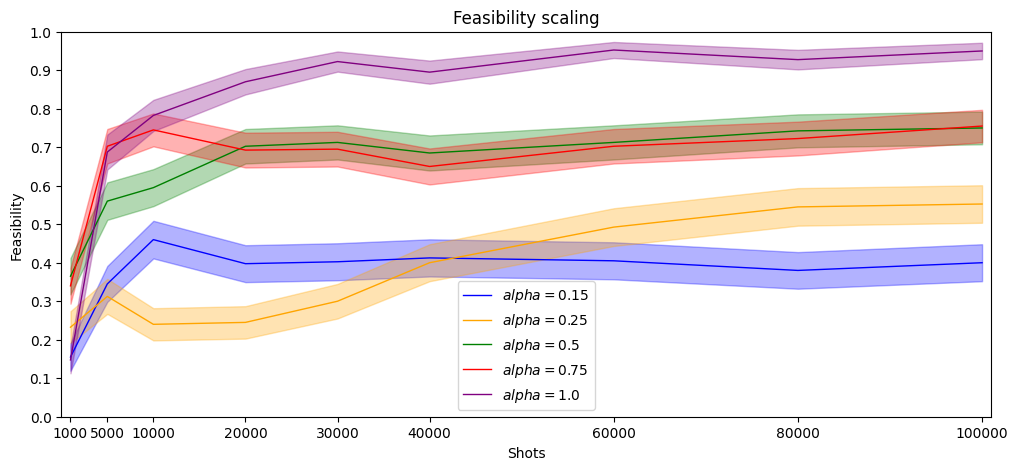}}
    \caption{Scaling of feasibility with $s$ for every cost function CVaR$_\alpha$ considered.}
    \label{fig:distr}
\end{figure} 

\subsubsection{Step 2 (Quality test)}
Based on the feasibility test, we have selected several VQAs (highlighted in red in Table~1) as candidates for well-suited algorithms. Next, we proceed to the quality test. Here, we establish a quality threshold of \(\mathcal{Q}_0 = 1.20\). Table~2 reports the measured VQA quality values, highlighting in blue those that pass the second test (provided they also passed the first one). As \(\alpha\) increases, the quality of the VQA tends to improve (see Figure~3). Again, we observe saturation in the scaling of quality with the number of shots, indicating that improvements beyond \(s = 20000\) are unclear for most cases, except for \(\alpha = 0.25\).
\begin{table}[!t]
\caption{Qualities of VQAs, for different $s$ and $\alpha$ values, with a confidence level of $95\%$.}
\centering
\renewcommand{\arraystretch}{1.3}
\setlength{\tabcolsep}{2pt}
\begin{tabular}{|>{\bfseries}c| *{5}{c|}}
\hline
\rowcolor{headercolor!50}
\boldmath $s\backslash\alpha$ & \bfseries 0.15 & \bfseries 0.25 & \bfseries 0.50 & \bfseries 0.75 & \bfseries 1.00 \\
\hline
\cellcolor{headercolor!50}\bfseries 1\,000 & 0.15 $\pm$ 0.04 & 0.25 $\pm$ 0.05 & 0.50 $\pm$ 0.07 & 0.53 $\pm$ 0.08 & 0.20 $\pm$ 0.05 \\
\cellcolor{headercolor!50}\bfseries 5\,000 & 0.34 $\pm$ 0.05 & 0.34 $\pm$ 0.05 & 0.75 $\pm$ 0.07 & \cellcolor{bluecell!20}{1.31 $\pm$ 0.09} & \cellcolor{darkred!20}{1.03 $\pm$ 0.08} \\
\cellcolor{headercolor!50}\bfseries 10\,000 & 0.45 $\pm$ 0.05 & 0.26 $\pm$ 0.05 & 0.80 $\pm$ 0.07 & \cellcolor{bluecell!20}{1.41 $\pm$ 0.09} & \cellcolor{bluecell!20}{1.20 $\pm$ 0.07} \\
\cellcolor{headercolor!50}\bfseries 20\,000 & 0.39 $\pm$ 0.05 & 0.26 $\pm$ 0.05 & \cellcolor{darkred!20}{0.94 $\pm$ 0.07} & \cellcolor{bluecell!20}{1.34 $\pm$ 0.09} & \cellcolor{bluecell!20}{1.32 $\pm$ 0.06} \\
\cellcolor{headercolor!50}\bfseries 30\,000 & 0.40 $\pm$ 0.05 & 0.32 $\pm$ 0.05 & \cellcolor{darkred!20}{0.96 $\pm$ 0.07} & \cellcolor{bluecell!20}{1.33 $\pm$ 0.09} & \cellcolor{bluecell!20}{1.36 $\pm$ 0.05} \\
\cellcolor{headercolor!50}\bfseries 40\,000 & 0.41 $\pm$ 0.05 & 0.42 $\pm$ 0.06 & \cellcolor{darkred!20}{0.92 $\pm$ 0.07} & {1.30 $\pm$ 0.10} & \cellcolor{bluecell!20}{1.32 $\pm$ 0.05} \\
\cellcolor{headercolor!50}\bfseries 60\,000 & 0.40 $\pm$ 0.05 & 0.52 $\pm$ 0.06 & \cellcolor{darkred!20}{0.97 $\pm$ 0.07} & \cellcolor{bluecell!20}{1.34 $\pm$ 0.09} & \cellcolor{bluecell!20}{1.40 $\pm$ 0.04} \\
\cellcolor{headercolor!50}\bfseries 80\,000 & 0.37 $\pm$ 0.05 & 0.58 $\pm$ 0.06 & \cellcolor{darkred!20}{1.01 $\pm$ 0.06} & \cellcolor{bluecell!20}{1.39 $\pm$ 0.09} & \cellcolor{bluecell!20}{1.34 $\pm$ 0.05} \\
\cellcolor{headercolor!50}\bfseries 100\,000 & 0.40 $\pm$ 0.05 & 0.59 $\pm$ 0.06 & \cellcolor{darkred!20}{1.02 $\pm$ 0.06} & \cellcolor{bluecell!20}{1.46 $\pm$ 0.09} & \cellcolor{bluecell!20}{1.38 $\pm$ 0.04} \\
\hline
\end{tabular}

\end{table}

\begin{figure}[!t]
    \centerline{\includegraphics[width=0.5\textwidth]{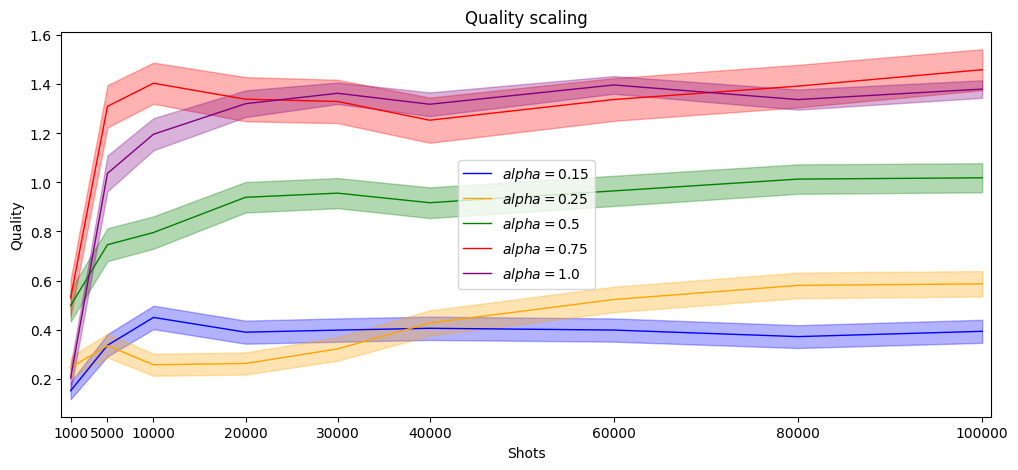}}
    \caption{Scaling of quality with $s$ for every cost function CVaR$_\alpha$ considered.}
    \label{fig:distr}
\end{figure}

\subsubsection{Step 3 (Reproducibility test)}
For the final test, we set the reproducibility threshold to $\mathcal{R}_0 = 0.60$. In Table~3, VQAs that pass the reproducibility test are highlighted in green. VQAs with lower values of $\alpha$ and $s$ have higher reproducibility, likely because the algorithm terminated without finding better solutions. Overall, Figure~4 shows that the number of shots does not significantly affect reproducibility. There is a slight upward trend for larger $\alpha$ values, but the effect remains nearly constant.
\begin{table}[!t]
\caption{Reproducibilities of VQAs, for different $s$ and $\alpha$ values, with a confidence level of $95\%$.}
\centering
\renewcommand{\arraystretch}{1.3}
\setlength{\tabcolsep}{2pt}
\begin{tabular}{|>{\bfseries}c| *{5}{c|}}
\hline
\rowcolor{headercolor!50}
\boldmath $s\backslash\alpha$ & \bfseries 0.15 & \bfseries 0.25 & \bfseries 0.50 & \bfseries 0.75 & \bfseries 1.00 \\
\hline
\cellcolor{headercolor!50}\bfseries 1\,000    & 0.69 $\pm$ 0.02 & 0.66 $\pm$ 0.03 & 0.54 $\pm$ 0.03 & 0.51 $\pm$ 0.03 & 0.61 $\pm$ 0.03 \\

\cellcolor{headercolor!50}\bfseries 5\,000    & 0.66 $\pm$ 0.02 & 0.68 $\pm$ 0.03 & {0.48 $\pm$ 0.03} &\cellcolor{bluecell!20} {0.55 $\pm$ 0.03} & \cellcolor{darkred!20}{0.32 $\pm$ 0.03} \\

\cellcolor{headercolor!50}\bfseries 10\,000   & 0.65 $\pm$ 0.02 & 0.72 $\pm$ 0.03 & 0.50 $\pm$ 0.03 &\cellcolor{greencell!20} {0.58 $\pm$ 0.03} & \cellcolor{bluecell!20}{0.36 $\pm$ 0.03} \\

\cellcolor{headercolor!50}\bfseries 20\,000   & 0.67 $\pm$ 0.02 & 0.69 $\pm$ 0.03 & \cellcolor{darkred!20}{0.48 $\pm$ 0.03} &\cellcolor{bluecell!20} {0.57 $\pm$ 0.03} & \cellcolor{bluecell!20}{0.42 $\pm$ 0.03} \\

\cellcolor{headercolor!50}\bfseries 30\,000   & 0.65 $\pm$ 0.02 & 0.67 $\pm$ 0.02 & \cellcolor{darkred!20}{0.49 $\pm$ 0.03} &\cellcolor{greencell!20} {0.60 $\pm$ 0.03} & \cellcolor{bluecell!20}{0.46 $\pm$ 0.04} \\

\cellcolor{headercolor!50}\bfseries 40\,000   & 0.66 $\pm$ 0.02 & 0.65 $\pm$ 0.02 & \cellcolor{darkred!20}{0.48 $\pm$ 0.02} & 0.57 $\pm$ 0.02 & \cellcolor{bluecell!20}{0.50 $\pm$ 0.04} \\

\cellcolor{headercolor!50}\bfseries 60\,000   & 0.68 $\pm$ 0.02 & 0.66 $\pm$ 0.02 & \cellcolor{darkred!20}{0.50 $\pm$ 0.03} & \cellcolor{greencell!20}{0.60 $\pm$ 0.03} & \cellcolor{greencell!20}{0.58 $\pm$ 0.04} \\

\cellcolor{headercolor!50}\bfseries 80\,000   & 0.67 $\pm$ 0.03 & 0.64 $\pm$ 0.02 & \cellcolor{darkred!20}{0.50 $\pm$ 0.03} & \cellcolor{greencell!20}{0.59 $\pm$ 0.03} & \cellcolor{greencell!20}{0.59 $\pm$ 0.03} \\

\cellcolor{headercolor!50}\bfseries 100\,000  & 0.64 $\pm$ 0.03 & 0.66 $\pm$ 0.02 & \cellcolor{darkred!20}{0.49 $\pm$ 0.03} & \cellcolor{greencell!20}{0.62 $\pm$ 0.03} & \cellcolor{greencell!20}{0.63 $\pm$ 0.03} \\
\hline
\end{tabular}

\end{table}

\begin{figure}[!t]
    \centerline{\includegraphics[width=0.5\textwidth]{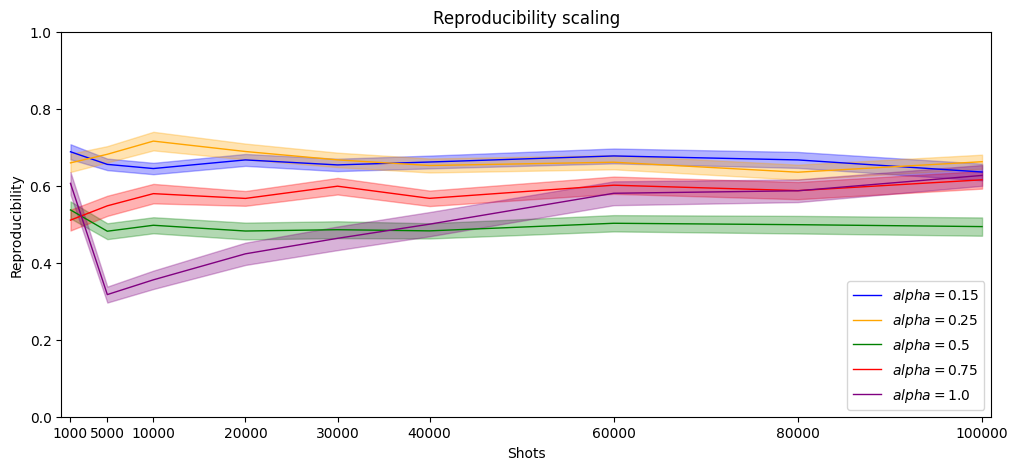}}
    \caption{Scaling of reproducibility with $s$ for every cost function CVaR$_\alpha$ considered.}
    \label{fig:distr}
\end{figure}
\noindent 

\subsubsection{Step 4 (Selection)}
From the previous tests, we conclude that the VQAs most suited to solve the defined QUBO problem, given the initial quantum circuit, are:
$\{ (0.75,10\,000),(0.75,30\,000),(0.75,60\,000),(0.75,80\,000),\\
   (0.75,100\,000),(1.00,60\,000),(1.00,80\,000),(1.00,100\,000)\}$.

This indicates that VQAs with higher values of $\alpha$ and $s$ are better suited to solve this problem given the initial quantum circuit.

\section{Discussion}
The proposed framework introduces a structured methodology for evaluating performance of VQAs applied to solve QUBO problems, addressing a critical gap in current benchmarking practices. Existing studies often report single-run performance metrics without explicitly accounting for the stochastic nature of VQA optimization. By incorporating feasibility, quality, and reproducibility into a unified evaluation process, our approach provides a more holistic view of algorithmic behavior under resource constraints.

\subsection{Theoretical Justification}
The framework is grounded in the observation that VQA performance is inherently stochastic due to sampling noise from finite measurement shots. Although classical optimizers such as COBYLA are deterministic, the optimization process becomes stochastic because the cost function is estimated from noisy samples of the quantum circuit's output distribution. This noise propagates through the optimization trajectory, leading to variability in the final solution even under identical parameter and hyperparameter settings. Our framework addresses this variability by introducing three complementary metrics:
\begin{itemize}
    \item \textbf{Feasibility} quantifies the probability of achieving a solution above a predefined success threshold, providing an a priori measure of algorithmic reliability.
    \item \textbf{Quality} formalizes the trade-off between resource consumption and success probability through a normalized distance in the quality diagram, enabling comparative efficiency analysis.
    \item \textbf{Reproducibility} leverages Shannon entropy to measure the concentration of outcomes in the quality diagram, offering a principled way to assess stability across repeated runs.
\end{itemize}

The use of entropy as a reproducibility metric is theoretically motivated by its ability to capture uncertainty in probability distributions. A concentrated distribution implies algorithmic robustness, while a high-entropy distribution signals sensitivity to noisy cost evaluations. Similarly, the quality diagram provides a geometric interpretation of optimization performance, where the ideal point $(0,0)$ represents maximum efficiency. These constructs collectively enable a multi-dimensional characterization of VQAs that goes beyond conventional single-objective evaluations, aligning with the need for systematic benchmarking in the NISQ era.

\subsection{Interpretation of Results}
Our proof-of-concept experiments reveal non-trivial relationships between resource allocation and performance. Feasibility and quality generally improve with the CVaR parameter $\alpha$ and the number of measurement shots $s$, confirming that increased sampling enhances success probability. However, reproducibility remains largely insensitive to these factors, suggesting that variability in outcomes is not fully explained through this performance metric.

The quality diagram introduced in this work provides an intuitive visualization of trade-offs between success probability and resource consumption. Beyond its descriptive role, this representation could serve as a foundation for adaptive strategies, where thresholds for feasibility, quality, and reproducibility are dynamically adjusted based on available computational budgets or application-specific requirements.

\subsection{Limitations}
Despite these contributions, the study has limitations. The analysis is restricted to a single QUBO instance, one ansatz, and one optimizer, all evaluated in simulation. While these choices demonstrate the framework’s applicability, they limit generalizability. Future work should explore broader problem classes, diverse ansätze, and real hardware implementations to validate the robustness of the proposed metrics under realistic noise conditions.

\section*{Conclusions \& Future Work}
We introduced a framework to evaluate the performance VQAs applied to solve QUBO problems, combining feasibility, quality, and reproducibility into a unified decision process supported by statistical estimation techniques. The framework provides a practical tool for comparing VQAs under resource constraints and contributes to more systematic evaluation practices for NISQ-era algorithms.

Future research should focus on extending the proposed framework beyond the proof-of-concept presented here. First, applying the methodology to a broader set of QUBO instances and other combinatorial optimization problems would help assess its robustness and generalizability. Additionally, exploring different ansätze—such as hardware-efficient circuits or QAOA—and a wider range of classical optimizers could provide insights into how these choices influence feasibility, quality, and reproducibility. Another important direction is validating the framework on real quantum hardware, where noise and connectivity constraints may significantly affect performance. Finally, adaptive selection criteria, such as dynamic thresholds or multi-objective optimization, could be investigated to improve decision-making under varying resource budgets. Integrating the framework into hybrid quantum-classical workflows for large-scale problems represents a promising avenue for practical deployment.

\bibliographystyle{unsrt}
\bibliography{references}

@article{glover2018tutorial,

  title={A Tutorial on Formulating and Using QUBO Models},

  author={Glover, Fred and Kochenberger, Gary and Du, Yang},

  journal={arXiv preprint arXiv:1811.11538},

  year={2018}

}

@book{garey1979computers,
  title={Computers and Intractability: A Guide to the Theory of NP-Completeness},
  author={Garey, M. R. and Johnson, D. S.},
  year={1979},
  publisher={W. H. Freeman}
}

@article{cerezo2021variational,
  title={Variational Quantum Algorithms},
  author={Cerezo, M. and Arrasmith, A. and Babbush, R. and Benjamin, S. C. and Endo, S. and Fujii, K. and ... Coles, P. J.},
  journal={Nature Reviews Physics},
  volume={3},
  pages={625--644},
  year={2021}
}

@article{barkoutsos2020improving,
  title={Improving Variational Quantum Optimization using CVaR},
  author={Barkoutsos, P. K. and Sokolov, I. O. and Moll, N. and Ganzhorn, M. and Salis, G. and Fuhrer, A. and Egger, D. J.},
  journal={Quantum},
  volume={4},
  pages={256},
  year={2020}
}

@article{ascendingcvar,
  title = {Evolving objective function for improved variational quantum optimization},
  author = {Kolotouros, Ioannis and Wallden, Petros},
  journal = {Phys. Rev. Res.},
  volume = {4},
  issue = {2},
  pages = {023225},
  numpages = {16},
  year = {2022},
  month = {Jun},
  publisher = {American Physical Society},
  doi = {10.1103/PhysRevResearch.4.023225},
  url = {https://link.aps.org/doi/10.1103/PhysRevResearch.4.023225}
}

@article{holland1975adaptation,
  title={Adaptation in Natural and Artificial Systems},
  author={Holland, J. H.},
  journal={University of Michigan Press},
  year={1975}
}

@article{kirkpatrick1983optimization,
  title={Optimization by Simulated Annealing},
  author={Kirkpatrick, S. and Gelatt, C. D. and Vecchi, M. P.},
  journal={Science},
  volume={220},
  number={4598},
  pages={671--680},
  year={1983}
}

@article{kochenberger2014unconstrained,
  title={Unconstrained Binary Quadratic Programming: A Survey},
  author={Kochenberger, G. A. and et al.},
  journal={Journal of Combinatorial Optimization},
  volume={28},
  pages={58--81},
  year={2014}
}

@article{farhi2014quantum,
  title={A Quantum Approximate Optimization Algorithm},
  author={Farhi, E. and Goldstone, J. and Gutmann, S.},
  journal={arXiv preprint arXiv:1411.4028},
  year={2014}
}

@article{perez2024variational,
  title={Variational Quantum Algorithms for Combinatorial Optimization},
  author={Daniel F. Perez-Ramirez},
  journal={arXiv preprint arXiv:2407.06421},
  year={2024}
}

@book{nielsen2002quantum,
  title={Quantum Computation and Quantum Information},
  author={Nielsen, M. A. and Chuang, I. L.},
  year={2002},
  publisher={Cambridge University Press}
}

@article{mcclean2016theory,
  title={The Theory of Variational Hybrid Quantum-Classical Algorithms},
  author={McClean, J. R. and Romero, J. and Babbush, R. and Aspuru-Guzik, A.},
  journal={New Journal of Physics},
  volume={18},
  number={2},
  pages={023023},
  year={2016}
}

@article{preskill2018quantum,
  title={Quantum Computing in the NISQ era and beyond},
  author={Preskill, J.},
  journal={Quantum},
  volume={2},
  pages={79},
  year={2018}
}

@article{bharti2022noisy,
  title = {Noisy intermediate-scale quantum (NISQ) algorithms},
  author = {Bharti, Kishor and Cervera-Lierta, Alba and Kyaw, Thi Ha and Haug, Tobias and Alperin-Lea, Samuel and Anand, Abhinav and Degroote, Matthias and Heimonen, Henri and Kottmann, Jakob S. and Menke, Tim and Mok, Wei-Keong and Sim, Sukin and Kwek, Leong-Chuan and Aspuru-Guzik, Al{\'a}n},
  journal = {Reviews of Modern Physics},
  volume = {94},
  number = {1},
  pages = {015004},
  year = {2022},
  doi = {10.1103/RevModPhys.94.015004}
}

@article{shannon1948mathematical,
  title = {A mathematical theory of communication},
  author = {Shannon, Claude E.},
  journal = {Bell System Technical Journal},
  volume = {27},
  number = {3},
  pages = {379--423},
  year = {1948},
  doi = {10.1002/j.1538-7305.1948.tb01338.x}
}

@article{bonet2021performance,
  title={Performance comparison of optimization methods on variational quantum algorithms},
  author={Bonet-Monroig, Xavier and Kyriienko, Oleksandr and Paternostro, Mauro and others},
  journal={arXiv preprint arXiv:2111.13454},
  year={2021}
}

@book{Casella2002,
  title={Statistical Inference},
  author={Casella, George and Berger, Roger L.},
  year={2002},
  publisher={Duxbury},
  edition={2nd}
}

@article{Paninski2003,
  title={Estimation of Entropy and Mutual Information},
  author={Paninski, Liam},
  journal={Neural Computation},
  volume={15},
  number={6},
  pages={1191--1253},
  year={2003},
  doi={10.1162/089976603321780272}
}

@article{Bickel1975,
  title={Some Contributions to the Theory of Statistical Estimation},
  author={Bickel, Peter J. and Doksum, Kjell A.},
  journal={Prentice Hall},
  year={1975}
}

@book{Taylor1997,
  title={An Introduction to Error Analysis: The Study of Uncertainties in Physical Measurements},
  author={Taylor, John R.},
  year={1997},
  publisher={University Science Books},
  edition={2nd}
}

@article{Basharin1959,
  title={On a statistical estimate for the entropy of a sequence of independent random variables},
  author={Basharin, G. P.},
  journal={Theory of Probability and its Applications},
  volume={4},
  number={3},
  pages={333--336},
  year={1959},
  doi={10.1137/1104035}
}

@article{schwagerl2024benchmarking,
  author = {Tim Schwägerl and Yahui Chai and Tobias Hartung and Karl Jansen and Stefan Kühn},
  title = {Benchmarking Variational Quantum Algorithms for Combinatorial Optimization in Practice},
  journal = {arXiv:2408.03073},
  year = {2024},
  url = {https://arxiv.org/abs/2408.03073}
}

@article{vinci2018optimally,
  author = {Walter Vinci and Alireza Shabani},
  title = {Optimally Stopped Variational Quantum Algorithms},
  journal = {Physical Review A},
  volume = {97},
  number = {4},
  pages = {042346},
  year = {2018},
  publisher = {American Physical Society},
  doi = {10.1103/PhysRevA.97.042346},
  url = {https://link.aps.org/doi/10.1103/PhysRevA.97.042346}
}

@book{CornuejolsTutuncu2006,
  author    = {Cornuéjols, Gérard and Tütüncü, Reha},
  title     = {Optimization Methods in Finance},
  publisher = {Cambridge University Press},
  year      = {2006},
  address   = {Cambridge, UK},
  note      = {Chapter 11: “Integer programming: theory and algorithms”},
}

@book{MansiniOgryczakSperanza2015,
  author    = {Mansini, Renata and Ogryczak, Włodzimierz and Speranza, M. Grazia},
  title     = {Linear and Mixed Integer Programming for Portfolio Optimization},
  publisher = {Springer},
  year      = {2015},
  address   = {Cham, Switzerland},
  isbn      = {978-3-319-18482-1},
}

@article{RenLuGuoXieZhangZhang2023,
  author    = {Ren, Yaping and Lu, Xinyu and Guo, Hongfei and Xie, Zhaokang and Zhang, Haoyang and Zhang, Chaoyong},
  title     = {A Review of Combinatorial Optimization Problems in Reverse Logistics and Remanufacturing for End-of-Life Products},
  journal   = {Mathematics},
  volume    = {11},
  number    = {2},
  pages     = {298},
  year      = {2023},
  doi       = {10.3390/math11020298},
}

@article{LouatiLahyaniAldaejMellouliNusir2021,
  author    = {Louati, Ali and Lahyani, Rahma and Aldaej, Abdulaziz and Mellouli, Racem and Nusir, Muneer},
  title     = {Mixed Integer Linear Programming Models to Solve a Real-Life Vehicle Routing Problem with Pickup and Delivery},
  journal   = {Applied Sciences},
  volume    = {11},
  number    = {20},
  pages     = {9551},
  year      = {2021},
  doi       = {10.3390/app11209551},
}

@article{blum1997selection,
  author  = {Avrim L. Blum and Pat Langley},
  title   = {Selection of Relevant Features and Examples in Machine Learning},
  journal = {Artificial Intelligence},
  volume  = {97},
  number  = {1-2},
  pages   = {245--271},
  year    = {1997},
  doi     = {10.1016/S0004-3702(97)00063-5},
}

@article{AloiseDeshpandeHansenPopat2009,
  author    = {Aloise, Daniel and Deshpande, Amit and Hansen, Pierre and Popat, Preyas},
  title     = {NP-hardness of Euclidean sum-of-squares clustering},
  journal   = {Machine Learning},
  volume    = {75},
  number    = {2},
  pages     = {245--248},
  year      = {2009},
  doi       = {10.1007/s10994-009-5103-0},
}

@article{HyafilRivest1976,
  author    = {Hyafil, L. and Rivest, R. L.},
  title     = {Constructing optimal binary decision trees is NP‐complete},
  journal   = {Information Processing Letters},
  volume    = {5},
  number    = {1},
  pages     = {15--17},
  year      = {1976},
  doi       = {10.1016/0020-0190(76)90095-8},
}

@book{PapadimitriouSteiglitz1998,
  author    = {Papadimitriou, Christos H. and Steiglitz, Kenneth},
  title     = {Combinatorial Optimization: Algorithms and Complexity},
  publisher = {Dover Publications},
  year      = {1998},
  address   = {Mineola, NY},
  edition   = {1},
}

@book{WolseyNemhauser1999,
  author    = {Wolsey, Laurence A. and Nemhauser, George L.},
  title     = {Integer and Combinatorial Optimization},
  publisher = {Wiley},
  year      = {1999},
  address   = {New York, NY},
  edition   = {1},
}

@article{Jones2024,
  author    = {Benjamin D. M. Jones and Lana Mineh and Ashley Montanaro},
  title     = {Benchmarking a Wide Range of Optimisers for Solving the Fermi-Hubbard Model Using the Variational Quantum Eigensolver},
  journal   = {arXiv preprint arXiv:2411.13742},
  year      = {2024},
}

@article{Zhu2024,
  author    = {Linghua Zhu and Senwei Liang and Chao Yang and Xiaosong Li},
  title     = {Optimizing Shot Assignment in Variational Quantum Eigensolver Measurement},
  journal   = {Journal of Chemical Theory and Computation},
  volume    = {20},
  number    = {6},
  pages     = {2390--2403},
  year      = {2024},
  doi       = {10.1021/acs.jctc.3c01113}
}

@article{du2022quantum,
  author = {Y. Du and others},
  title = {Quantum Circuit Architecture Search for Variational Quantum Algorithms},
  journal = {npj Quantum Information},
  volume = {8},
  article = {1},
  year = {2022},
  doi = {10.1038/s41534-022-00570-y}
}

@article{wu2024variational,
  author  = {Wu, Kai and others},
  title   = {Variational Benchmarks for Quantum Many-Body Problems},
  journal = {Science},
  year    = {2024},
  doi     = {10.1126/science.adg9774}
}

@book{kushner2003stochastic,
  author    = {Harold J. Kushner and G. George Yin},
  title     = {Stochastic Approximation and Recursive Algorithms and Applications},
  publisher = {Springer},
  year      = {2003},
  edition   = {2nd}
}

@book{spall2003introduction,
  author    = {James C. Spall},
  title     = {Introduction to Stochastic Search and Optimization: Estimation, Simulation, and Control},
  publisher = {Wiley},
  year      = {2003}
}

@article{he2022,
  author       = {Guang Ping He},
  title        = {Training quantum machine learning models on cloud without uploading the data},
  journal      = {arXiv preprint arXiv:2206.08453},
  year         = {2022},
  url          = {https://arxiv.org/abs/2206.08453}
}

@article{abbas2021,
  author       = {Abbas, Amira and Sutter, David and Zoufal, Christa and Lucchi, Aurelien and Figalli, Alessio and Woerner, Stefan},
  title        = {The power of quantum neural networks},
  journal      = {Nature Computational Science},
  volume       = {1},
  pages        = {403--409},
  year         = {2021},
  doi          = {10.1038/s43588-021-00084-1}
}

@article{arthur2022,
  author       = {Arthur, D. and Date, P.},
  title        = {A hybrid quantum-classical neural network architecture for binary classification},
  journal      = {arXiv preprint arXiv:2201.01820},
  year         = {2022},
  url          = {http://arxiv.org/abs/2201.01820}
}

@incollection{powell1994cobyla,
  author       = {Powell, M. J. D.},
  title        = {A Direct Search Optimization Method That Models the Objective and Constraint Functions by Linear Interpolation},
  booktitle    = {Advances in Optimization and Numerical Analysis},
  editor       = {Gomez, S. and Hennart, J.-P.},
  series       = {Mathematics and Its Applications},
  volume       = {275},
  pages        = {51--67},
  publisher    = {Kluwer Academic},
  address      = {Dordrecht},
  year         = {1994},
  doi          = {10.1007/978-94-015-8330-5_4}
}

@article{Bishwas2024PortfolioQA,
  title={Comparative analysis of diverse methodologies for portfolio optimization leveraging quantum annealing techniques},
  author={Tang, Zhijie and Dou, Alex Lu and Bishwas, Arit Kumar},
  journal={arXiv preprint arXiv:2403.02599},
  year={2024},
  url={https://arxiv.org/abs/2403.02599}
}

@article{Bishwas2024ParallelQA,
  title={Investigation into the Potential of Parallel Quantum Annealing for Simultaneous Optimization of Multiple Problems: A Comprehensive Study},
  author={Bishwas, Arit Kumar and Som, Anuraj and Choudhary, Saurabh},
  journal={arXiv preprint arXiv:2403.05764},
  year={2024},
  url={https://arxiv.org/abs/2403.05764}
}

@article{Bishwas2024MolecularUnfolding,
  title={Molecular unfolding formulation with enhanced quantum annealing approach},
  author={Bishwas, Arit Kumar and Pitchai, Arish and Som, Anuraj},
  journal={arXiv preprint arXiv:2403.00507},
  year={2024},
  url={https://arxiv.org/abs/2403.00507}
}

\appendices

\section{Statistical consistency}
\label{app:qubo-matrix}
\paragraph{Feasibility estimation}
Using the central limit theorem for the binomial distribution \cite{Casella2002}, the sample size required to estimate feasibility at a given significance level, $S.L.$ is:
\begin{equation}
    n = \frac{z_{S.L./2}^2\,p(1-p)}{E^2},
\end{equation}
where $z_{S.L./2}$ is the z-score corresponding to the desired confidence level, $p$ denotes the feasibility, and $E$ represents half the width of the confidence interval. Setting an error of $5\%$ with a confidence level of $95\%$, we obtain approximately $n = 384$ optimizations in the worst-case scenario ($p = 0.5$), which we round up to $n = 400$. Feasibility is then computed by verifying whether the final circuit is feasible in each run and calculating the proportion of feasible circuits across the $n = 400$ repetitions.

\paragraph{Quality estimation}
The quality of a VQA is defined as the expected value of a cost function over the probability distribution of its outcomes. Applying the central limit theorem, we estimate it as:
\begin{equation}
    \hat{\mathcal{Q}} = \frac{1}{n} \sum_{i=1}^{n} q_i,
\end{equation}
with standard deviation given by:
\begin{equation}
    \hat{\sigma}_\mathcal{Q} = \sqrt{\frac{1}{n-1} \sum_{i=1}^n (q_i - \hat{\mathcal{Q}})^2},
\end{equation}
and the corresponding error computed as:
\begin{equation}
    \Delta \mathcal{Q} = z_{S.L./2} \frac{\hat{\sigma}_\mathcal{Q}}{\sqrt{n}},
\end{equation}
where $z_{S.L./2}$ is the z-score associated with the desired confidence level.

\paragraph{Reproducibility estimation}
To quantify reproducibility, we compute the Shannon entropy of the distribution:
\begin{equation}
    S = -\sum_{i=1}^K p_i \ln(p_i),
\end{equation}
where $K$ denotes the number of bins. The quality diagram is discretized into grids of size $0.1 \times 0.1$, resulting in $K = 100$. The normalized reproducibility metric is then given by:
\begin{equation}
    \hat{\mathcal{R}} = 1 + \frac{1}{\ln K} \sum_{i=1}^K p_i \ln(p_i).
\end{equation}
The entropy estimator is asymptotically normal \cite{Paninski2003,Basharin1959}:
\begin{equation}
    \sqrt{n} (\hat{S} - S) \sim \mathcal{N}(0, \sigma_S^2), \quad n \rightarrow \infty,
\end{equation}
where
\begin{equation}
    \sigma_S^2 = \mathrm{Var}[\ln(p)].
\end{equation}

The delta method \cite{Casella2002,Bickel1975} yields a consistent estimator for the variance:
\begin{equation}
    \hat{\sigma}_S^2 = \frac{1}{n} \left( \sum_{i=1}^K \hat{p}_i \ln^2(\hat{p}_i) - \left( \sum_{i=1}^K \hat{p}_i \ln(\hat{p}_i) \right)^2 \right).
\end{equation}
Thus, the error in the entropy estimate is:
\begin{equation}
    \Delta \hat{S} = z_{S.L./2} \sqrt{\hat{\sigma}_S^2},
\end{equation}
which propagates to an error in reproducibility using standard error propagation \cite{Taylor1997}:
\begin{equation}
    \Delta \mathcal{R} = \frac{1}{\ln K} \Delta \hat{S}.
\end{equation}

\end{document}